\documentclass[conference]{IEEEtran}
\IEEEoverridecommandlockouts
\usepackage{cite}
\usepackage{amsmath,amssymb,amsfonts}
\usepackage{algorithmic}
\usepackage{graphicx}
\usepackage{tabularx}
\usepackage{booktabs}
\usepackage{textcomp}
\usepackage{xcolor}
\usepackage{url}
\def\BibTeX{{\rm B\kern-.05em{\sc i\kern-.025em b}\kern-.08em
    T\kern-.1667em\lower.7ex\hbox{E}\kern-.125emX}}
\begin{document}

%\title{The Perception of Translational Shape-Changing Haptic Interfaces Varies with Stimulus Magnitude, Direction, and Grasp Type\\
\title{Investigating the Perception of Translational Shape-Changing Haptic Interfaces\\
% {\footnotesize \textsuperscript{*}Note: Sub-titles are not captured for https://ieeexplore.ieee.org  and
}

\author{\IEEEauthorblockN{Qihan Yang*}
\IEEEauthorblockA{Imperial College London\\London, UK\\qihan.yang21@imperial.ac.uk}
\and
\IEEEauthorblockN{Xin Zhou*\thanks{All authors are from the Manipulation and Touch Lab (MTL), Department of Electrical and Electronic Engineering, Imperial College London, London, UK. (*Qihan Yang and Xin Zhou are co-first authors.)}}
\IEEEauthorblockA{Imperial College London\\London, UK\\xin.zhou16@imperial.ac.uk}
\and
\IEEEauthorblockN{Adam J. Spiers}
\IEEEauthorblockA{Imperial College London\\London, UK\\a.spiers@imperial.ac.uk}
}

\vspace{-2mm}
\maketitle

\begin{abstract}
Shape-changing haptic interfaces (SCHIs) are a promising and emerging field. However, compared to more established stimulus modalities, such as vibration, there is sparse literature on the perception of dynamic shapes. Furthermore, the influence of properties such as grasp types and displacement magnitude/direction has not been formally evaluated. This work attempts to initiate a formal perceptual evaluation of SCHIs via a psychophysical user study involving a 1-DOF translational shape-changing interface that can move its body with 1.25-micrometer resolution. Participants completed a \textit{Method of Constant Stimulus} study while holding the device with three different grasps. Stimuli direction occurred both toward and away from the thumb, while the standard stimuli varied between small (0.48\,mm) and large (6\,mm). Our results indicate that translational SCHIs should maximize the translation magnitude rather than the number of fingers in contact. We also demonstrated how to apply our findings to real-world applications via a simple `paddle game', where we compared conventional linear mapping with non-linear mapping derived from our perceptual experiment outcomes between the device position and its represented value.
%Based on the outcomes of the perceptual experiments, we also implemented a simple ‘paddle’ game to compare the linear and non-linear mapping between the device position and its represented value.
Results indicate that the non-linear mapping was more effective, with improved error distribution. We hope this work inspires further formal perceptual investigation into other SCHI morphologies. 
\end{abstract}


\section{Introduction}
Shape-changing haptic interfaces (SCHIs) are an emerging area of HCI research that promises highly intuitive data communication with low-cognitive load \cite{rasmussen2012shape,alexander2018grand}. Though SCHIs are an exciting area of development, the majority of touch researchers and device designers still rely on classical haptic feedback modalities. As an example, the Oct-Dec 2024 issue of \textit{IEEE Transactions on Haptics} featured 46 papers, of which 15 were focused on vibration feedback, while only two involved shape change. 

We conjecture that current limited research into SCHIs may be due to several factors. One is the lack of readily available hardware platforms in comparison to vibration actuators (which are inexpensive and easy to interface). Even open-source shape-changing devices must be 3D printed and assembled with various fasteners and actuators (e.g. \cite{Spiers-SBAN}). Another factor is that shape-change has severely limited perceptual resources in the literature compared to other haptic modalities, such as vibrotactile or audio feedback. %This lack of understanding of how users perceive shape-change stimuli confounds SCHI development.

\begin{figure}[t]
  \includegraphics[width=\columnwidth]{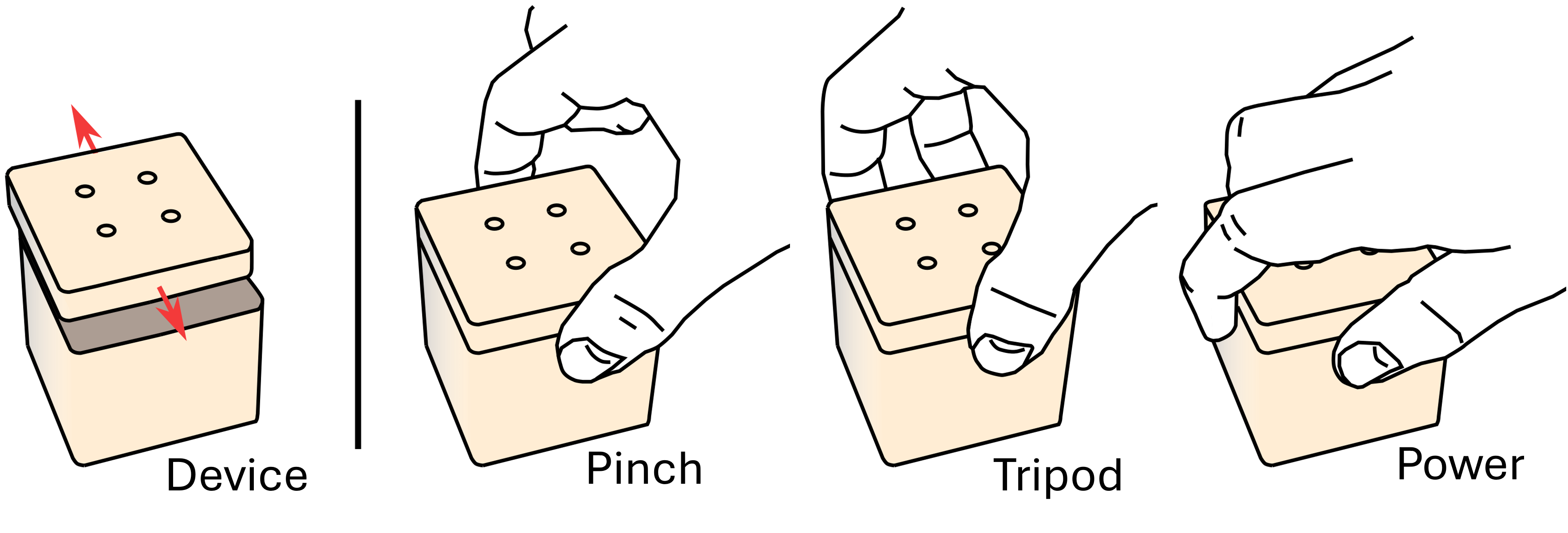}
  \caption{The shape-changing haptic interface and three grasp types during the experiment. Pinch utilises only the index finger and thumb. Tripod includes the middle finger. Power involves four digits and the palm.}
  \label{fig:grasp_types}
  \vspace{-3mm}
\end{figure}

\begin{figure*}[t]
\centering
  \includegraphics[width=0.98\textwidth]{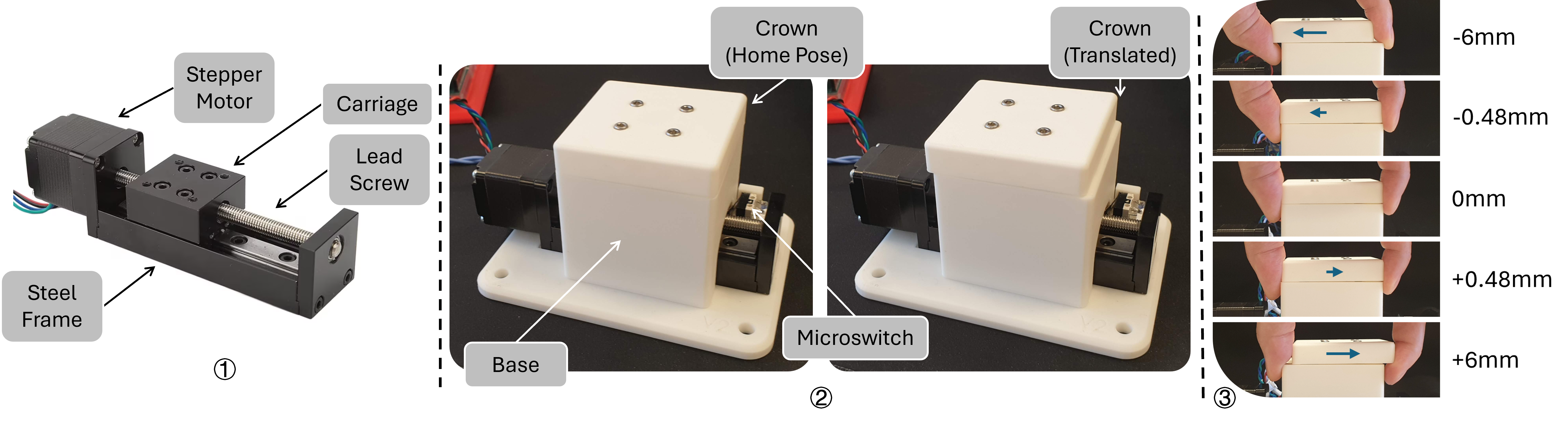}
  \caption{The test rig is based on a high-precision linear actuator secured in a 3D-printed housing. As the sliding interface reaches greater displacement, finger contacts become more complex, affecting users' perception of change.}
  \label{fig:Testrig}
   \vspace{-3mm}
\end{figure*}

% Our short term goal is to aid understanding of the translational form of SCHI and aid future implementations and design changes of such devices. Our longer-term goal is to initiate a comprehensive investigation into the perception of diverse SCHI morphologies for a better understanding of the whole field. 
In this work, we present the first in-depth psychophysical study of an SCHI. We implement a cube-shaped translational test rig, whose two aligned faces become misaligned through lateral translation (Figures~\ref{fig:grasp_types} and \ref{fig:Testrig}). Via a psychophysical study using classical methods, we investigate how the following variables affect user perception:

\begin{enumerate}
    \item Translation magnitude (`small' and `large')
    \item Translation direction (away from and toward the thumb)
    \item Grasp type (pinch, tripod and power)
\end{enumerate}

Furthermore, we apply the results of the perceptual study to an objective task by implementing a simple `paddle' video game. Participants are asked to `catch' invisible falling balls, whose lateral positions are conveyed via our test rig. We use the game platform to compare two mappings between the ball position on the screen and the shape-changing response: a linear mapping and a non-linear mapping based on the psychophysical results.

Our short term goal is to aid understanding of the translational form of SCHI and aid future implementations and design changes of such devices. Our longer-term goal is to initiate a comprehensive investigation into the perception of diverse SCHI morphologies (such as those mentioned in \cite{yoshida2020pocopo, ryu2021gamesbond}) to enable better understanding across the whole field. 
% We intend our empirical findings to set the scene for further psychophysical studies on other SCHIs with different morphologies, such as those mentioned in \cite{yoshida2020pocopo, ryu2021gamesbond} and help the future SCHIs' design.

\section{Related Work and Research Gap}
% \begin{figure*}[t]
% \centering
%   \includegraphics[width=0.3\textwidth]{figures/Annotated3.png}
%   \includegraphics[width=0.2\textwidth]{figures/Finger Offset V1.png}
%   \includegraphics[width=0.4\textwidth]{figures/User Study.png}
%   \caption{The test rig is based on a high-precision linear actuator (intended for use in a CNC machine) secured in a 3D-printed housing. As the sliding interface reaches greater displacement, finger contacts become more complex, affecting change perception.}
%   \label{fig:Testrig}
% \end{figure*}
Shape-changing haptic interfaces have diverse designs and serve different functions. Some SCHIs are desk-mounted and can render different shapes, softness, and/or texture \cite{follmer2013inform,boem2019human,Zhu2022,Suzuki2021}. Other systems have been designed to be handheld and portable \cite{ryu2021gamesbond,yoshida2020pocopo,jang2016haptic,Spiers-SBAN,hemmert2010shape,spiers2016design,kim2016fast, yechendeshi, Quinn2024}.
Despite many SCHIs being developed, there exists little literature that addresses how users perceive changing shapes. This is particularly true when compared to other sensory modalities, such as vibration or sound. The psychophysical evaluation of vibration has been a topic of study for many decades (e.g. \cite{verrillo1985psychophysics} from 1985), with more recent comparisons being made between the perception of vibration and sound \cite{merchel2020psychophysical}. Such works have inspired further psychophysical research on the perceptual qualities of other haptic modalities, including skin-stretch \cite{Schorr_SkinStretch,Provancher_SkinStretch}, air-jets \cite{gwilliam2012characterization} and magneto-rheological brakes \cite{chen2024development}. The outcomes of such studies give interface designers valuable insights needed to anticipate a user's perception of stimuli prior to developing an interface or application. Notably, detailed information of this type is currently absent for SCHIs.

To encourage further SCHI development, we conducted a psychophysical experiment studying the user perception of `translational' shape change. Translation, in which a part slides from the main body, is involved in several SCHIs, especially in the navigation tool group, where the distance and orientation of a target to the user are often indicated by the magnitude and direction of translation\cite{Spiers-SBAN,hemmert2010shape,spiers2016design,yechendeshi}. These two translational feedback parameters (magnitude and direction), as well as how the users grasp the device, vary across previous publications~\cite{ryu2021gamesbond,jang2016haptic}. Therefore, the scope of this work is studying the effect of different translation magnitudes, directions, and grasp types on the perception of translational haptic shape change.

\section{Hardware Design}
% \begin{figure}[t]
% \centering
%   \includegraphics[width=\columnwidth]{figures/testrig.png}
%   \caption{The test rig is based on a high-precision linear actuator secured in a 3D-printed housing. As the sliding interface reaches greater displacement, finger contacts become more complex, affecting change perception.}
%   \label{fig:Testrig}
% \end{figure}
To facilitate the psychophysics study, we built the hardware interface illustrated in Figure \ref{fig:Testrig}. This test rig is partly inspired by the cube-shaped \textit{Animotus} \cite{spiers2016design,Spiers2018Theatre} and \textit{Deshi} devices \cite{yechendeshi}. Notice that the device is a prototype desktop-based test-bed specifically for perceptual studies. It has a simplified 1DOF output, high positional accuracy and high force capability with a non-backdrivable transmission.

The test rig is based around a high-precision linear actuator (manufactured by \textit{Fafeicy}), intended for use in CNC machines. The actuator consists of a stepper motor connected to a lead screw which moves a carriage along a linear bearing. The actuator has a stroke length of 50\,mm and a minimum step size of 0.00125\,mm with our stepper driver. We mounted the steel frame onto a 3D-printed platform, which was connected to the base of our cube-shaped device. A crown was then attached to the actuator's carriage, enabling translational motion. All 3D-printed components were printed in PLA using a Raise3D E2 printer. A microswitch was mounted on the platform to enable automatic homing of the crown, which was carried out before every experiment. The stepper motor was controlled via a \textit{Goodn TB6600} stepper motor driver connected to an Arduino Uno, which received serial commands from a PC running the Python experiment software.

\section{Perceptual Study Methods}
We hypothesize that users’ perception of a translational shape-changing device depends on grasp type, translation magnitude, and translation direction. To test this, we conducted a series of psychophysics experiments using Gescheider’s \textit{Methods of Constant Stimulation – Difference Thresholds} \cite{gescheider1985psychophysics} to measure and compare perception across these factors.

\begin{figure}[t]
  \centering
  \includegraphics[width=0.95\columnwidth]{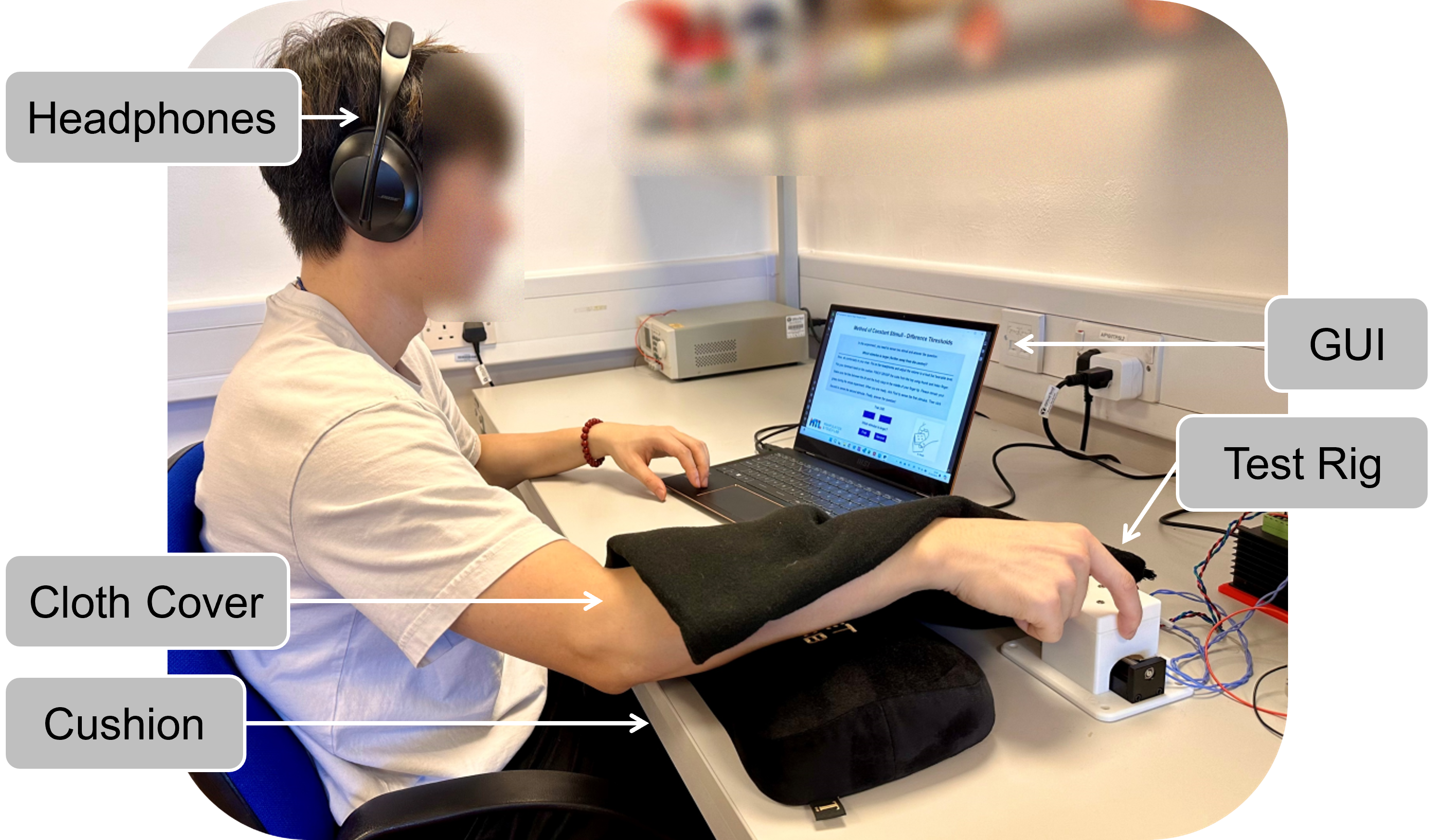}
  \caption{User study setup and GUI. The cloth cover has been partially removed for this photograph.}
  \label{fig:User Study}
  \vspace{-3mm}
\end{figure}

Throughout the experiment, participants maintained a consistent grasp on the device with their dominant hand, ensuring continuous contact with both the device' base and crown. They were instructed not to apply excessive force that could impede the crown's translational movement. Headphones (playing white noise) and a thick cloth (over the hand and device) reduced audio and visual cues (Figure~\ref{fig:User Study}). Participants were presented with pairs of stimuli following a two-alternative, forced-choice (2AFC) paradigm, where the order of standard and comparison stimuli was randomized. Each stimulus began with the device’s crown centered on the base (home pose), then translated either away from or toward the thumb to a pre-defined magnitude, and then back to the home pose. After both stimuli were presented, participants indicated which one (first or second) appeared greater (further away from the center).

Within each combination of chosen factors (grasp type, magnitude, and direction), the standard stimulus was fixed at the chosen magnitude. The comparison stimulus then used magnitudes spaced symmetrically around the standard stimulus (in increments of a fixed step size)~\cite{gescheider1985psychophysics}.

\subsection{Choosing Experiment Parameters}
% \subsubsection{Grasp Type}
Based on the Feix grasp taxonomy ~\cite{Feix2016}, three major grasp types were selected: Pinch, Tripod, and Power (Figure \ref{fig:grasp_types}).
% \subsubsection{Stimulus Direction and Magnitude}
Per grasp, we investigated two translation magnitudes (a small and a large standard stimulus) in both directions (away from or toward the thumb), symmetric around the centered position. 

A series of short 2AFC pilot experiments were conducted with two pilot participants. During these sessions, the standard stimuli magnitudes and the step sizes were heuristically adjusted. The smaller standard stimuli needed to be close to zero yet far away for the comparison stimuli to not `overflow' in the opposite direction. The larger standard stimuli had to be far enough from zero to elicit complex finger contact patterns (Figure \ref{fig:Testrig}.3), yet close enough to hold comfortably. Comparison magnitudes were also chosen adequately to construct clear psychophysical curves to reliably quantify perceptions (with details on quantification methods given in the next section).
% A series of short 2AFC pilot experiments were conducted with two pilot participants, during which the magnitudes were heuristically adjusted, 
This process resulted in the following stimulus levels:
\begin{itemize}
    \item Smaller ±0.48\,mm standard, with 5 comparison at ±0.8, ±0.64, ±0.48, ±0.32, and ±0.16\,mm (0.16\,mm steps);
    \item Larger ±6\,mm standard, with 5 comparison at ±8, ±7, ±6, ±5, ±4\,mm (1\,mm steps).
\end{itemize}

A third pilot participant took part in an extended-length perceptual experiment (1.5 hours with 120 comparisons) to verify our selection. Note that different comparison step sizes had to be chosen for the two standard magnitudes as the perception performances varied significantly. None of the pilot participants took part in the main experiment.

%Though it is unconventional to vary step size, it was necessary to get clear S-shaped psychometric curves for the two standard stimuli.

\subsection{Main Perceptual Experiment}
Ten participants took part in the main experiment (four female, six male, mean age = 25.3). For each participant, the experiment was split into three rounds, one for each grasp type. In every round, the four standard stimuli were compared against the five corresponding comparison stimuli four times each ($4\times5\times4=80$). This resulted in $3\times80=240$ comparison trials per participant, taking around one hour. In total, we collected $10\times240=2400$ trials. The order of grasp type, comparison pairs within rounds, and the crown's moving speed were all balanced, predetermined, and pseudo-randomized to reduce bias. Participants received gift vouchers as compensation for their time. This study was approved by the Ethics committee of Imperial College London (application number 6640164).

%Imperial College Science, Engineering, Technology, Research  Ethics Committee (application number 6640164).

\section{Perceptual Study Analysis and Results}

% Through the experiments, we obtained enough data to compare 3 grasp types, 2 directions per grasp type, and 2 magnitudes per direction. 

Through the experiments, we obtained enough data to compare two stimulus magnitudes, three grasp types per magnitude, and two stimulus directions per magnitude. 

The perceptual data was processed in Python using the \textit{Python-Psignifit} toolbox (\url{https://github.com/wichmann-lab/python-psignifit}) to fit psychometric functions. Three measures were used for the analysis: 1) Point of Subjective Equality (PSE) is defined as the stimulus level that is perceived to be larger than a given standard stimulus 50\% of the time (or at the 50\% confidence threshold). 2) Just Noticeable Difference (JND) is the stimulus level difference between the 25\% and 75\% confidence threshold. It conveys how large of a discrepancy in stimulus magnitude is needed for the user to notice the difference. 3) Weber Fraction (WF) \cite{ADictionaryofPsychology} calculates the JND in proportion to the corresponding standard stimulus level and is a measure of how sensitive our sensory system is to percentage changes of that stimulus magnitude. 

%Note that many stimuli types (e.g. vibrotactile, weight, light intensity) have predominantly constant Weber Fractions across stimuli levels (Weber's Law \cite{ADictionaryofPsychology}). 
%However, interacting with shape-changing devices is more complex. For example, at small translations, a user's hand pose stays largely constant, and shape-change is perceived simply via skin deformation, whereas at larger translations, the hand pose is forced to adapt to the device shape with a much more complex contact shape. In this work, the Weber Fraction is used as a measure of sensitivity, with smaller values indicating a higher sensitivity.

\begin{figure*}[t]
  \centering
  \includegraphics[width=\textwidth]{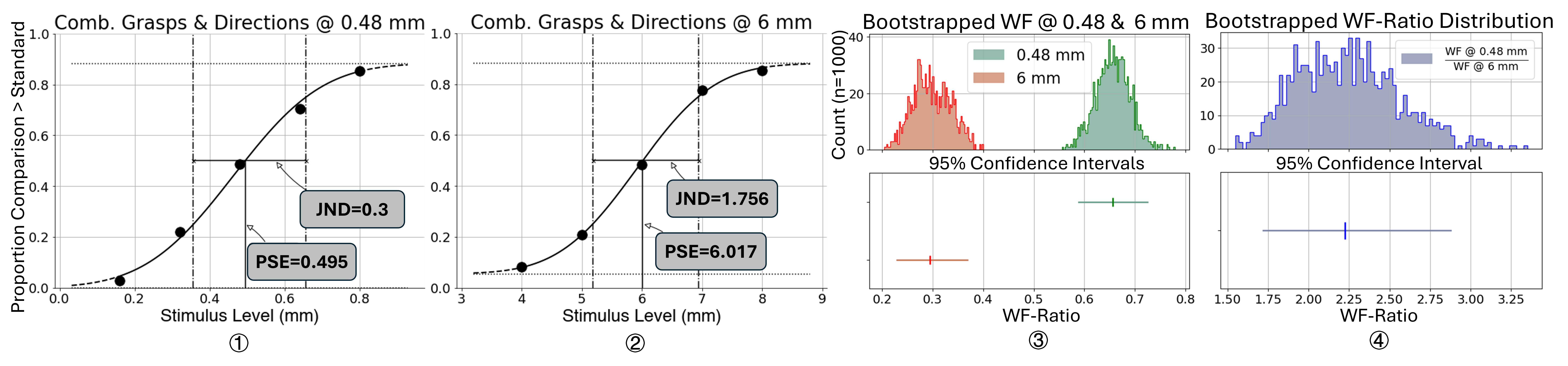}
  \vspace{-8mm}
  \caption{1 and 2: Psychometric plots with all grasps combined and no differentiation in direction. The PSE and JND are labelled. 3: Distribution of 1000 bootstrapped WF at two stimulus levels (0.48\,mm and 6\,mm). 4: Distribution of 1000 bootstrapped paired WF-ratios. Individual WF and WF-ratio's 95\% Confidence Intervals (CIs) are shown with the median marked. 
  % For WF-ratio, the lower and upper CIs are 1.72 and 2.88.
  }
  \label{fig:all}
  \vspace{-3mm}
\end{figure*}

\subsection{Bootstrapped WF-ratio}
We could not confidently assess the normality of the WF distributions if individually fitted to each of the 10 participants. Hence, we opted for a robust non-parametric analysis method based on bootstrapping to compare the user perception between conditions (two stimulus magnitudes, three grasp types, or two stimulus directions). 

First, ten random participants are selected with replacement (meaning that the same participant can be selected more than once). Their collective responses under one condition (e.g. Tripod) are used to fit an s-curve to calculate a $WF_1$. The collective responses from the same pool of participants under a second condition (e.g. Pinch) are then used to calculate a second $WF_2$. The ratio of these paired WFs $WF_1/WF_2$ (e.g. Tripod/Pinch) is also calculated. These steps are repeated 1000 times, getting 1000 WF-ratios and forming a large bootstrapped WF-ratio distribution. 

This method uses a ten times larger response pool to fit each s-curve, producing smaller fitting errors than using individual participant responses. The statistical significance can be assessed by determining whether the confidence intervals of the bootstrapped WF-ratio distributions cross the value 1. This method also provides a tangible way to quantify potential improvements from one condition to another.

\subsection{Results by Stimulus Magnitude}
\label{sec:mag}
\begin{table}[b]
  \vspace{-3mm}
  \centering
  \caption{Psychometric Results by Stimulus Magnitude}
  \begin{tabularx}{\columnwidth}{@{}XXXX@{}}
%     \toprule
% & \multicolumn{2}{c}{\textbf{Psychometric Results by Stimulus Magnitude}} \\ 
    % \cmidrule(r){2-4}
    \toprule
    {\small Standard Stimuli (mm)}& 
    {\small Collective PSE (mm)}&
    {\small Collective JND (mm)}&
    {\small Collective WF}
    \\
    \midrule
    0.48& 0.495& 0.300& 0.625 \\
    6& 6.017& 1.756& 0.293\\
    \bottomrule
  \end{tabularx}
  \label{tab:magnitude}
\end{table}

The results in this section focus on comparing perception at the two chosen standard stimulus magnitudes, 0.48\,mm and 6\,mm, and do not distinguish between grasp types nor directions (e.g., +6\,mm and -6\,mm are both treated as 6\,mm). Figures \ref{fig:all}.1 and \ref{fig:all}.2 show the psychophysical curves fitted on the combined (non-bootstrapped) responses of the 10 participants. Table \ref{tab:magnitude} lists the collective PSEs, JNDs, and WFs (from the same non-bootstrapped curves). Figures \ref{fig:all}.3 and \ref{fig:all}.4 show the distribution of the two standard stimulus bootstrapped WFs along with the distribution of the WF-ratios (0.48\,mm and 6\,mm). 

Despite the larger JND of 1.756\,mm at 6\,mm magnitude, the WF (0.293) is only half that of the WF (0.625) at 0.48\,mm. Figure \ref{fig:all}.4 shows that the distribution of WF-ratios well exceeds 1, indicating a statistically significant difference between the WFs from the two stimulus levels. From the 95\% confidence interval (CI), the WF at 0.48\,mm is between 1.72 and 2.88 times as large as at 6\,mm. This indicates that a large translation magnitude corresponds to a significantly heightened perception sensitivity. Note that this does not follow Weber's Law \cite{ADictionaryofPsychology}, which states that most sensory modalities (e.g. vibration, weight, light intensity) have predominantly constant Weber Fractions across stimuli levels. We conjecture that this is because a large translation displacement results in more complex contact shapes (shown in Figure~\ref{fig:Testrig}), which provide additional information to users.

\subsection{Results by Grasp Type}
\begin{table}[b]
\vspace{-3mm}
  \centering
  \caption{Psychophysical Results By Grasp Types }
  \begin{tabularx}{\columnwidth}{@{}XXXXX@{}}
%     
% & \multicolumn{3}{c}{\textbf{Psychometric Results by Grasp Type}} \\ 
    \toprule
& \multicolumn{3}{c}{\small{\textbf{0.48\,mm Standard Stimulus}}} \\
\midrule
    {\small\textit{Grasp Type}}
    & {\small \textit{Collective PSE (mm)}}
    & {\small \textit{Collective JND (mm)}}
    & {\small \textit{Collective WF}}
    & {\small \textit{Average WF}} \\
    \midrule
    Pinch & 0.461& 0.219& {0.456} & 0.566\\
    Tripod & 0.532& 0.294& 0.612 & {0.558}\\
    Power & 0.485& 0.321& 0.669 & 0.680\\
    \toprule
    %\end{tabularx}
    %\begin{tabularx}{\columnwidth}{@{}XXXXX@{}}
    %     \toprule
    & \multicolumn{3}{c}{\small{\textbf{6\,mm Standard Stimulus}}} \\
        \midrule
        Pinch & 5.846 & 1.486 & {0.248} & {0.239} \\
        Tripod & 6.241 & 2.117 & 0.353 &0.334 \\
        Power & 6.068 & 2.108 & 0.351 &0.298 \\
         \bottomrule
      \end{tabularx}
      \label{tab:psychophysical outcomes grasp}
\end{table}

\label{sec:grasp}
\begin{figure*}[t]
  \centering
  \includegraphics[width=\textwidth]{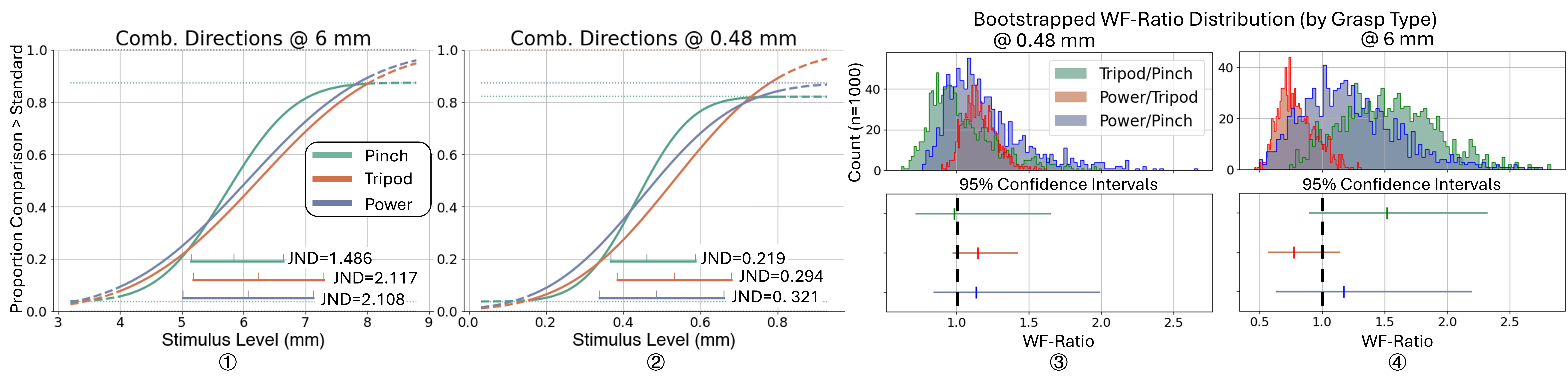}
  \vspace{-8mm}
  \caption{1 and 2: Overlaid perceptual plots for the three grasp types. The bottom bars in each plot mark the 25\%, 50\% (PSE), and 75\% percentiles for each grasp type. The JND is the length of each bar. 3 and 4: Distribution of 1000 bootstrapped paired WF-ratios between the three grasp types (Tripod/Pinch, Power/Tripod, and Power/Pinch) at two standard stimulus levels. 95\% confidence intervals are shown in the bottom half with the median marked.}
  \label{fig:Overlay_Results}
  \vspace{-3mm}
\end{figure*}

This section compares the three grasp types without distinguishing stimulus direction. Comparisons were conducted separately for ±0.48\,mm and ±6\,mm magnitudes, as responses from these levels cannot be combined. The overlaid psychometric plots of the three grasp types are shown in Figures \ref{fig:Overlay_Results}.1 and \ref{fig:Overlay_Results}.2. Table \ref{tab:psychophysical outcomes grasp} presents the collective PSEs, JNDs, WFs, and Average WF (calculated via fitting each participant individually). Looking at the collective WF, the Pinch grasp resulted in much lower (better) WF at both 0.48\,mm (0.456) and 6\,mm (0.248) standard stimuli levels compared to the other two grasp types.

% \begin{figure}[t]
%   \centering
%   \includegraphics[width=0.85\columnwidth]{figures/Bootstrap_by_grasp.png}
%   \caption{Distribution of 1000 pair-wise bootstrapped WF-ratios between the Pinch, Tripod, and Power grasps at two stimulus levels. 95\% confidence intervals are shown in the bottom half with the median marked.}
%   \label{fig:boot_grasps}
%   %\vspace{-10pt}
% \end{figure} 

Figures \ref{fig:Overlay_Results}.3 and \ref{fig:Overlay_Results}.4 show the distributions of bootstrapped WF-ratios between the three grasp types in a pair-wise manner (`more contact fingers'/`fewer contact fingers'). This comparison did not show any statistical significance, since all the 95\% CIs cross 1. However, we do observe the majority of WF-ratios to exceed 1.
%via these CIs, we are able to find some interesting insights. 
For 6\,mm standard stimulus, the WF-ratio of Tripod/Pinch has a lower CI bound of 0.90 and a median above 1 (1.5). The WF-ratio of Power/Pinch has a lower CI bound of 0.63 and also a median above 1 (1.17). This means that the Tripod's and Power's WF is at most 10.0\% and 37.0\% lower (better) than Pinch's WF, respectively. Similarly, at the 0.48\,mm magnitude, Tripod's WF is at most 28.0\% better than Pinch, while Power's WF is at most 2.4\% and 15.31\% better than Tripod and Pinch, respectively. Combining these results, we conclude that our results show a nonsignificant trend toward better performance for fewer contact fingers, and that choosing a grasp with more fingers can at most marginally improve perception, if not make it worse. A small number of participants commented that they focused on one single finger most of the time, which could explain these results. %This trend warrants further investigation with a larger sample size to determine if the observed effect is reliable.

\subsection{Results by Direction}
\begin{figure}[t]
  \centering
  \includegraphics[width=\columnwidth]{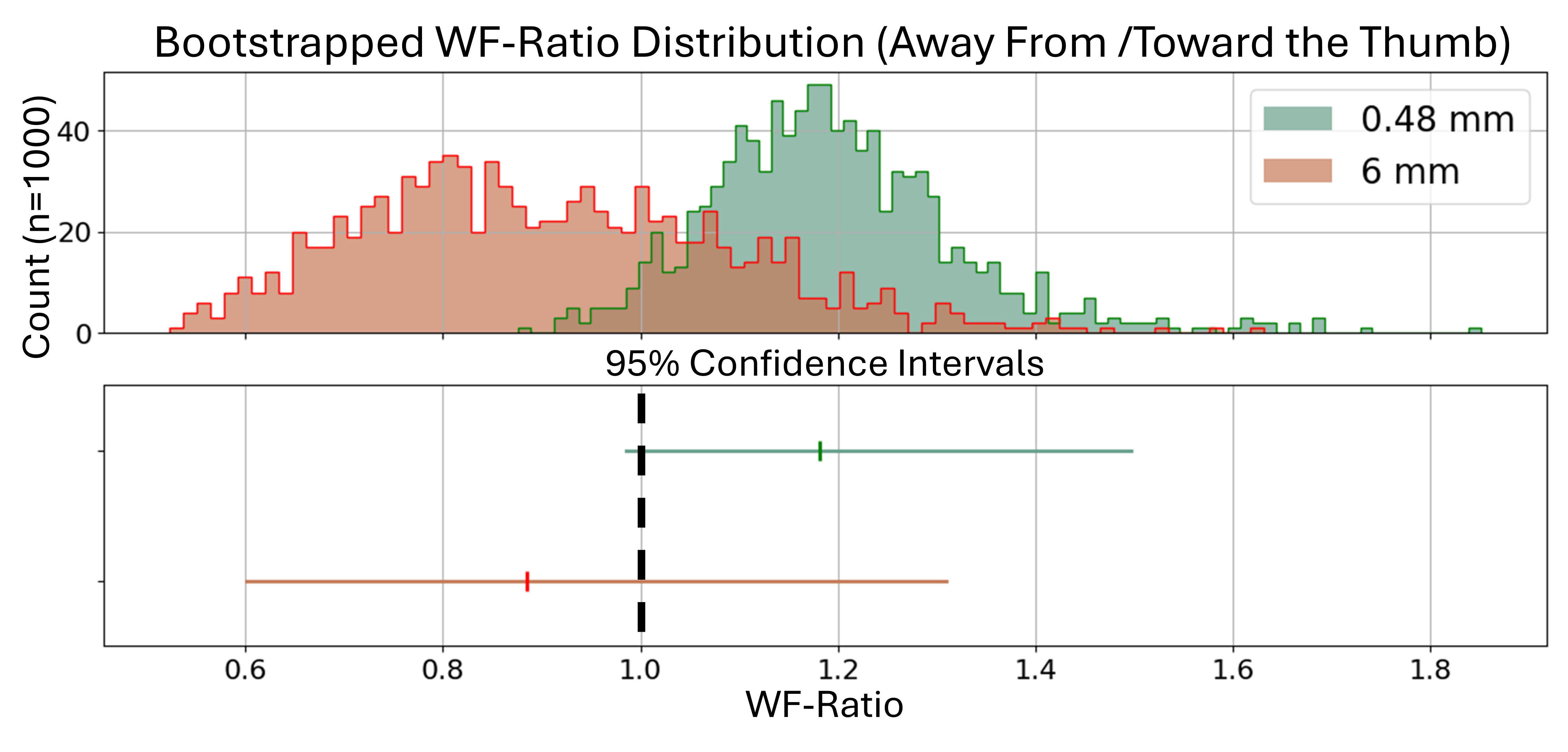}
  \vspace{-3mm}
  \caption{Distribution of 1000 bootstrapped paired WF-ratios (WFs of `away from thumb divided by WFs of `toward thumb') at two stimulus levels. 95\% confidence intervals are shown in the bottom half with the median marked.}
  \label{fig:boot_dir}
  \vspace{-3mm}
\end{figure}
\label{sec:dir}

The results in this subsection are separated by stimulus direction with the grasps combined. As before, responses from the small and large magnitudes cannot be combined, so the effects of stimulus direction are compared separately under those two magnitudes. Figure \ref{fig:boot_dir} shows the distributions of the bootstrapped WF-ratios between translation away from and toward the thumb. No statistical significance between the two translation directions is observed for either magnitude level. However, for the small magnitude, the majority of WF-ratios above 1 implies a nonsignificant trend toward better performance (smaller potential downside) for moving toward the thumb. For the large magnitude, the WF-ratios are widespread with no notable trend.

After experiments, three participants reported (unprompted) that they noticed a clear difference between translation directions. We conjecture that the directional bias exists, likely due to the inherent asymmetry of human hands. However, we could not infer a universally preferred direction from the statistic, probably because the directional bias differs between people and can exhibit different magnitudes in different directions~\cite{Voudouris2016, Romano2019}.

\section{Paddle Game: Applying our Findings}
\subsection{Paddle Game Methods}
After the psychophysical study, we wanted to demonstrate how our findings can be applied for improving user interactions with SCHIs. We implemented a simple game loosely based on the 1976 video game \textit{Breakout} \cite{wiki:Breakout_(video_game)} using the Unity game engine. In our game, the user controls the horizontal position of an on-screen paddle (using a laptop touchpad) to catch 100 falling balls. This game is a simple 1D analogy to real-world applications, such as screen reader cursor position representation for vision-impaired users.

\begin{figure}[t]
% \vspace{-0.3cm}
  \centering
  \includegraphics[width=\columnwidth]{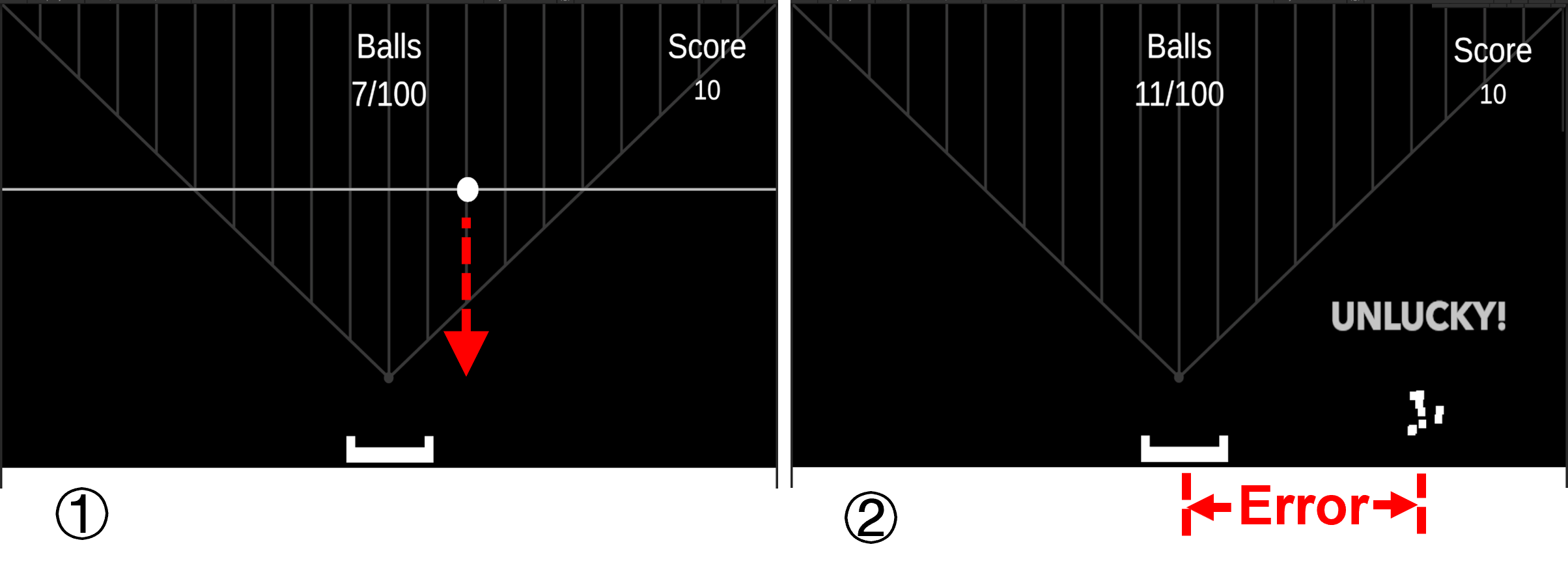}
  \vspace{-6mm}
   \caption{Screenshots from the paddle game. 1: One of the ten visible balls used for training. 2: Invisible balls. A message of `Perfect’, `Not Bad’ or `Unlucky’ was shown depending on whether and how well the ball was caught. The red arrows, lines and text are for annotation purposes only and were not visible to participants.}
   \label{fig:Paddle Game}
   \vspace{-5mm}
\end{figure}

During the game, the first 10 balls were visible to help the user calibrate their perception (Figure \ref{fig:Paddle Game}.1), and the remaining 90 balls were invisible (Figure \ref{fig:Paddle Game}.2). The balls' vertical positions were represented via a falling horizontal line and the balls' horizontal positions were only represented to the user through the position of the shape-changing device. A ball falling in the middle of the screen is presented as the device in its home pose (0\,mm in Figure \ref{fig:Testrig}). A ball falling in the left/right half of the screen is presented as the device translates from its home pose to the left/right to a certain magnitude (based on a mapping scheme), then back to its home pose. The ball positions were presented in a pre-generated pseudo-random order and ensured to be evenly distributed across the screen space and game time. For the 90 invisible balls, the pixel distances between their position when hitting the `ground' and the paddle center were averaged as the error (Figure \ref{fig:Paddle Game}.2). 

\begin{figure*}[t]
  \centering
  \includegraphics[width=0.9\textwidth]{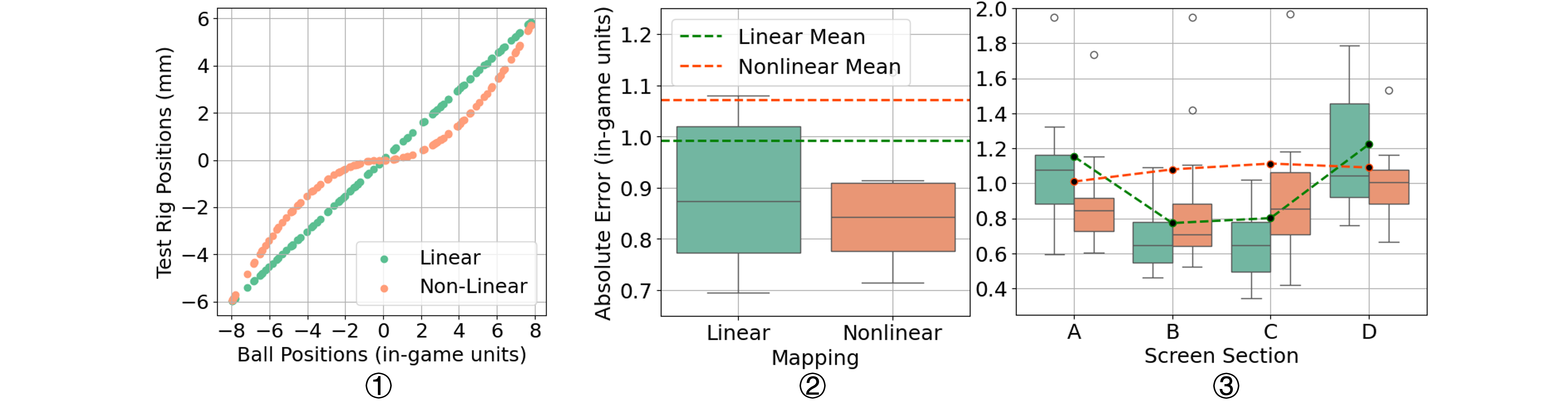}
  \vspace{-4mm}
   \caption{1: the linear versus nonlinear mapping. Each dot represents a ball position and its corresponding device position. 2: Overall errors under the two mapping schemes. Each box plot contains 16 points, each being the error of an individual participant playing under a given mapping scheme. 3: Sectional errors by splitting the balls into four screen regions. From left, A: Balls in the first 25\% of the screen. B: Balls in the 25\% to 50\% range. C: Balls in the 50\% to 75\% range. D: Balls in the last 25\% region. For 2 and 3, the extreme outliers have been cropped from the view.}
   ~\label{fig:Game}
   \vspace{-3mm}
\end{figure*}

The following psychophysical outcomes were applied during game design. Firstly, the participants used Pinch grasp with their dominant hand when playing the game. Secondly, the JND was taken into consideration. In the context of our game, the JND represents the required device translation for the user to perceive a change in ball position. Based on the fact that the JND increases with the translation magnitude, we hypothesize that a simple linear mapping between the ball's horizontal position ($x$) and device position (translation magnitude $y$) would lead to lower errors toward the center but higher errors near the screen edges. We then implemented an improved non-linear mapping derived from our psychophysical results and expected to find more evenly spread errors. Both schemes map -8 to 8 game units (ball position) to -6\,mm to 6\,mm on the device (Figure \ref{fig:Game}.1) and are listed below:
\begin{itemize}
    \item Linear Mapping: $y = \frac{3}{4}x$
    \item Non-linear Mapping: $y = sgn(x)\frac{3}{32} x^{2}$
\end{itemize}

For the non-linear mapping, the gradient $\frac{dy}{dx}$, representing the JND, approximately follows the values and trend observed in the perception study: when $y = \pm0.4$, $\frac{dy}{dx} = 0.39$, and when $y = \pm6$, $\frac{dy}{dx} = 1.5$. This gradient is significantly smaller near the screen center $x=0$ compared to the screen edges $x=\pm8$.

\subsection{Paddle Game Experiment and Results}

16 participants took part in the game, 6 of which were returning volunteers from the previous perception study. In total, 12 were male, 3 were female, and one participant preferred not to disclose their gender (average age 27.19). As a result, there were 16 paired participant-wise average absolute errors for the two mapping schemes, shown in Figure \ref{fig:Game}.2 as box plots. The overall mean errors are close (0.991 for linear mapping and 1.071 for non-linear mapping). There is no statistical difference between the two groups based on a non-parametric distribution-free paired Wilcoxon Signed-Rank Test~\cite{Rey2011} (\textit{p} = 0.980, Effect size \textit{r} = -0.013, Cohen's \textit{d} = -0.230). This statistical information conveys that participants' overall performances on both mapping methods are similar.

However, as we aimed to even out the errors across the screen with our non-linear mapping, we split the ball-position range into four equal horizontal sections and consolidated the equal amount of balls falling within each section (each section also contains 16 paired average absolute errors). Figure \ref{fig:Game}.3 presents the box plots and the mean value of the four groups of sectional paired errors. Under the linear mapping scheme, the section error distribution is largely uneven, with larger errors concentrated toward the edges of the screen. With the perception-based non-mapping function, on the other hand, a much more uniform error distribution is present with smaller errors for the two sections closer to the screen edge. The medium to large effect sizes of sections A, C, and D show that the differences between the two mapping methods are statistically meaningful (\textit{p} = [A: 0.083, B: 0.464, C: 0.004, D: 0.144], Effect size \textit{r} = [A: -0.44, B: -0.194, C: -0.685, D: -0.375], Cohen's \textit{d}: [A: 0.442, B: -0.419, C: -0.715, D: 0.41]). These results imply that our non-linear mapping helps the users have an even perception across a large translation range. Another interesting finding is that the errors on the left half are smaller than those on the right. Considering the majority of our participants are right-handed (14 out of 16), this implies a slightly better perception of translation toward the thumb, which aligns with our trend found in Section~\ref{sec:dir}.
%For right-handed subjects (14/16 in our study, the majority), this implies a slightly better perception of translation toward the thumb, which aligns with our slight bias found in Section~\ref{sec:dir}.

\section{Conclusion}
This work investigated several factors that can impact user perception of translational shape change: grasp type, translation magnitude, and translation direction. The results imply, firstly, that at larger shape-changing magnitudes (which result in more complex contacts), the perception sensitivity is significantly better than at smaller magnitudes. This improvement in sensitivity (WF) over magnitude could possibly be a unique characteristic of translational shape-changing devices that are not found in common haptic modalities which have a somewhat constant sensitivity. Secondly, having more fingers in contact with a shape-changing device does not effectively result in better perception. Thirdly, the perception of shape-changing devices is not symmetric and can have a directional bias, which is likely to be different from person to person. Finally, we demonstrated how our findings can influence the design of real-world haptic devices and applications via a paddle game. We derived a non-linear mapping scheme based on the JNDs from our psychophysics studies and found that it indeed managed to achieve a more evenly distributed perception of on-screen positions. Our methodology can be extended far beyond the paddle game and improve various translational haptic guidance systems for both on-screen and off-screen tasks. 

We hope that our findings can act as a road map for this emerging area of SCHIs design in various ways. For example, a compact ping-pong ball-sized shape-changing device pinched between two fingertips with a dynamic shape-change gradient might just outperform a baseball-sized linear shape-changing device held within the palm. As a natural next step, we plan on creating and evaluating a more optimized translational shape-changing device in addition to expanding our perceptual analysis of other forms of shape change (such as twisting, expansion, or bending).

% \section{Acknowledgments}
% We thank all volunteers who (very patiently) participated in these studies. 

\bibliographystyle{IEEEtran}
\bibliography{ref}

\end{document}